\documentclass[aps,prl,twocolumn,showpacs]{revtex4}
\usepackage{graphicx}
\usepackage{amsmath}

\begin{document}

\title{An Alternative Interpretation of Recent ARPES Measurements on TiSe$_2$}

\author{Jasper van Wezel}
  \affiliation{Cavendish Laboratory, University of Cambridge, Madingley Road, Cambridge CB3 0HE, UK}
\author{Paul Nahai-Williamson}
  \affiliation{Cavendish Laboratory, University of Cambridge, Madingley Road, Cambridge CB3 0HE, UK}
\author{Siddarth S. Saxena}
  \affiliation{Cavendish Laboratory, University of Cambridge, Madingley Road, Cambridge CB3 0HE, UK}

\begin{abstract}
Recently there has been a renewed interest in the charge density wave transition of TiSe$_2$, fuelled by the possibility that this transition may be driven by the formation of an excitonic insulator or even an excitonic condensate. We show here that the recent ARPES measurements on TiSe$_2$ can also be interpreted in terms of an alternative scenario, in which the transition is due to a combination of Jahn-Teller effects and exciton formation. The hybrid exciton-phonons which cause the CDW formation interpolate between a purely structural and a purely electronic type of transition. Above the transition temperature, the electron-phonon coupling becomes ineffective but a finite mean-field density of excitons remains and gives rise to the observed diffuse ARPES signals.
\end{abstract}


\maketitle

\subsection*{Introduction}
The nature of the Charge Density Wave (CDW) transition observed in TiSe$_2$ at 202 K had been the subject of intense debate for more than three decades \cite{LiCDW, Wilson, Zunger, Cercellier, Whangbo, Suzuki, Motizuki, Hughes}. During most of this time, the three main hypotheses competing over the explanation of the driving force behind the CDW phase were the formation of an excitonic insulator \cite{LiCDW, Wilson, Zunger, Cercellier}, the presence of an indirect Jahn-Teller effect \cite{Whangbo, Suzuki, Motizuki}, and the involvement of a band-type Jahn-Teller effect \cite{Hughes}. The more usual picture of Fermi surface nesting has been ruled out \cite{Rossnagel}. Some experimental support has been found for both of the Jahn-Teller effects as well as for the Excitonic Insulator scenario, but there is no conclusive evidence for the dominance of any one of them. Very recently, a series of high-resolution angle-resolved photo-emission spectroscopy (ARPES) measurements have been interpreted as showing evidence that the CDW transition in TiSe$_2$ may be driven by the formation of an excitonic condensate, rather than by any of the previously mentioned scenarios \cite{Monney, Monney1, Monney2}. In this article, we suggest that these recent ARPES measurements can also be interpreted in terms of a specific combination  of the Jahn-Teller and Excitonic Insulator scenarios which we recently introduced to both unite the existing experimental results on TiSe$_2$ at ambient pressure and to describe the superconducting phase which appears upon the suppression of CDW order at high pressures \cite{vanWezel,vanWezel1,Anna}.

We first give a brief account of the different existing hypotheses and the experimental evidence for them, before reviewing the recent ARPES results and their interpretation in terms of an excitonic condensate. We then comment on the experimental and theoretical implications of this scenario, and suggest an alternative interpretation in which the CDW formation is driven by a combination of excitons and phonons. 

\subsection*{Jahn-Teller and Excitonic Insulator Scenarios}
In its normal phase (at high temperatures), TiSe$_2$ consists of hexagonal layers of Ti sandwiched by octahedrally coordinated Se atoms, the so-called 1T-polytype for transition metal dichalcogenides.It is either a semimetal \cite{DiSalvo, LiCDW,  Wilson} or a semiconductor \cite{Kidd, Rossnagel,Rasch} with a small indirect gap of order 100 meV or less. At 202 K it undergoes a transition into a commensurate 2x2x2 CDW state \cite{DiSalvo}. The driving mechanism behind this CDW formation has been suggested to be some variant of the Jahn-Teller effect in which a commensurate spatial reconstruction of the lattice lowers the average energy of both the conduction and the valence bands close to the Fermi surface by facilitating partial charge transfer between neighbouring 4$p$ and 3$d$ orbitals. This could happen through either a lowering of the local bonding state energy, the so-called indirect Jahn-Teller effect \cite{Whangbo}, or by causing an alteration in the local crystal field around the Ti atoms, the band-type Jahn-Teller effect \cite{Hughes}. The main support for this scenario comes from the agreement between the experimentally observed commensurate distortion pattern with the instabilities due to these Jahn-Teller effects as seen in various electronic band structure calculations \cite{Holt,Suzuki}. It is unclear however how to fully account for the large observed transfer of spectral weight to the backfolded bands in the CDW phase within a scenario driven purely by Jahn-Teller effects \cite{Cercellier, Monney1, Voit}. 

The competing hypothesis that the formation of an Excitonic Insulator underlies the CDW transition \cite{Wilson}, naturally gives rise to large amounts of spectral weight redistribution. The exciton formation is made possible by the paucity of charge carriers in the system and the correspondingly poorly screened Coulomb interaction. With sufficient electron-hole coupling between the valence and conduction bands, the system is unstable to the formation of excitons and deforms with a periodicity governed by the wave-vector connecting them. Because of the poor nesting of electron and hole pockets in the Fermi surfaces of TiSe$_2$ however, it is not clear how the electron-hole coupling alone could explain the emergence of a commensurate rather than an incommensurate CDW.

\subsection*{The Combined Scenario}
\begin{figure}[t]
\centerline{{\includegraphics[width=1.0 \linewidth]{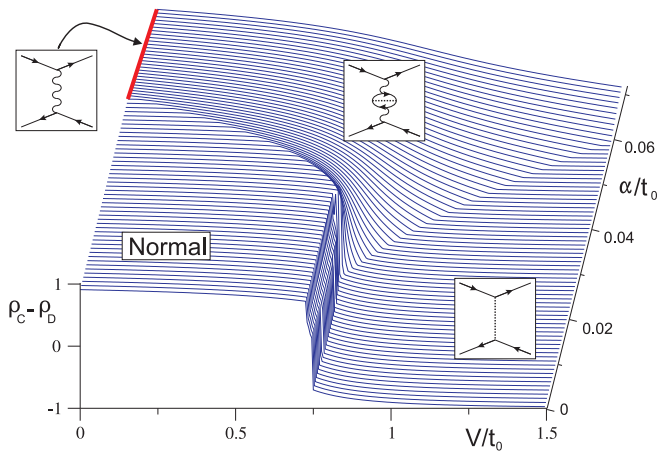}}}
\caption{(Color online) The zero-temperature charge transfer $\rho_C-\rho_D$ in the mean field treatment of equation \eqref{Hamiltonian}, as a function of the exciton binding energy $V$ and the electron-phonon coupling strength $\alpha$ (both normalised with respect to the unperturbed bandwidth $t_0$). The Feynman diagrams indicate the different instabilities of the normal (low $V$, low $\alpha$) state. From left to right, they are driven purely by electron-phonon coupling, by a mixture of phonons and excitons, and purely by the excitonic binding energy.}
\label{LineDiagramFinal}
\end{figure}

We have recently shown that the strengths of both of these approaches can be combined into a single scenario in which exciton formation and electron-phonon coupling (i.e. Jahn-Teller effects) cooperate to form a CDW phase characterised by both a large amount of charge transfer and the observed commensurate distortion pattern \cite{vanWezel}. We constructed a quasi one-dimensional model based on a tight-binding description of TiSe$_2$ which shows how the balance between electron-phonon coupling and exciton binding energy can lead to competition, coexistence and even cooperation in the formation of an ordered (CDW) phase. The Hamiltonian for the model can be written as:
\begin{align}
\hat{H}=\hat{H}_0 + \sum_{\langle i,j \rangle} \alpha \hat{\varphi}_{ij} \hat{c}^{\dagger}_i \hat{d}^{\phantom \dagger}_j - V \hat{d}^{\dagger}_i \hat{d}^{\phantom \dagger}_i \left( 1 - \hat{c}^{\dagger}_j \hat{c}^{\phantom \dagger}_j \right) + \text{H.c.}
\label{Hamiltonian}
\end{align}
Here $\hat{H}_0$ includes the unperturbed band structure of the conduction and valence bands with parameters determined by a tight binding fit to LDA calculations \cite{vanWezel,vanWezel1}. Electrons in the valence band are created by the $\hat{c}^{\dagger}$ operators while $\hat{d}^{\dagger}$ creates a conduction band electron and $\hat{\varphi}_{ij}$ displaces the ions along the bond connecting nearest neighbour sites $i$ and $j$ on the quasi one-dimensional lattice defined in ref. \cite{vanWezel}. The different phases found in this model system are shown in figure \ref{LineDiagramFinal}. 

These phases can be distinguished by looking at the difference between electronic density in the valence and the conduction band ($\rho_C-\rho_D=\langle c^{\dagger}c-d^{\dagger}d \rangle$) as a function of the excitonic binding energy $V$. In the region of low binding energy and low electron-phonon coupling the system is in its normal state and $\rho_C-\rho_D$ is (nearly) maximal and constant. Upon increasing $V$, an excitonic instability may be reached and at zero temperature all available electrons become bound into electron-hole pairs, thus creating an excitonic insulator. The transition into this phase is sharp and characterised by the sudden inversion of $\rho_C-\rho_D$. Alternatively, one can increase the electron-hole coupling parameter $\alpha$ while keeping the excitonic binding energy equal to zero. In this case a structural instability is encountered which displaces the atomic positions, but does not directly lead to any charge transfer. The order parameter for this phase is of the form $\tau=\langle c^{\dagger}d+d^{\dagger}c \rangle$. The transition thus gives rise to as a kink in the evolution of $\rho_C-\rho_D$ rather than a direct discontinuity. For $\alpha$ and $V$ both finite, an intermediate type of transition can occur in which an increase in $V$ leads to the renormalisation and softening of the phonon spectrum which in turn causes the structural instability. In this case both $\tau$ and $\rho_C-\rho_D$ form part of the order parameter and the CDW is mediated by hybridised exciton-phonons.

\subsection*{ARPES Measurements and the Condensate Scenario}
Recently it was suggested that the Excitonic Insulator scenario could be extended to an Excitonic Condensate scenario, in which the CDW transition is driven not merely by the formation of bound particle-hole pairs, but by their condensation into a phase-coherent BCS-like state \cite{Monney, Monney1}. The original motivation for this assertion was the excellent agreement between a set of high-definition ARPES measurements of the band structure close to the Fermi energy and a theoretical model band structure based on the assumption that a BCS-like mechanism would open up a gap in the single particle excitation spectrum. Unfortunately the gap itself was taken to a be a free fitting parameter in these calculations, making it impossible to draw any definitive conclusions about the origin of the gap. More specifically, one could start from a model which includes only electron-phonon coupling and no excitons, and find that upon driving the model through a structural instability to open up a gap, the functional form of the resulting Green's functions in terms of the size of the gap is precisely equal to the ones considered in ref. \cite{Monney1}.

Additional motivation for the condensate scenario was presented very recently, in the form of temperature dependent ARPES measurements which show that at high temperatures a diffuse remnant of the low-temperature backfolded valence band remains visible below the conduction band \cite{Monney2}. This signal was interpreted as evidence for the presence of strong, incoherent electron-hole pair fluctuations.

The Excitonic Condensate scenario and the Excitonic Insulator model share a difficulty in accounting for the observed periodicity of the distortion pattern in TiSe$_2$. In the absence of a coupling to the lattice, the lack of a nesting vector connecting the electron and hole pockets of the Fermi surface makes the emergence of a commensurate CDW by purely electronic means unlikely. In addition, the Excitonic Condensate scenario needs to account for the difference between the lattice responses induced by the high temperature, incoherent excitonic fluctuations on the one hand and the low temperature, coherent excitonic condensate on the other.

A further theoretical complication comes from the general observation that the formation of a phase-coherent excitonic condensate on a lattice is severely hindered by the presence of Umklapp-processes \cite{Guseinov,Littlewood2}. In the presence of only Coulomb interactions, these processes explicitly break the U(1) phase symmetry and preclude the possibility of having a true phase transition into a condensed phase \cite{Littlewood2}. The transition into an Excitonic Insulator phase will be rendered first order by the same processes, as opposed to the second order transition observed in TiSe$_2$ \cite{Guseinov}.

\subsection*{An Alternative Interpretation}
\begin{figure}[t]
\centerline{{\includegraphics[width=1.0 \linewidth]{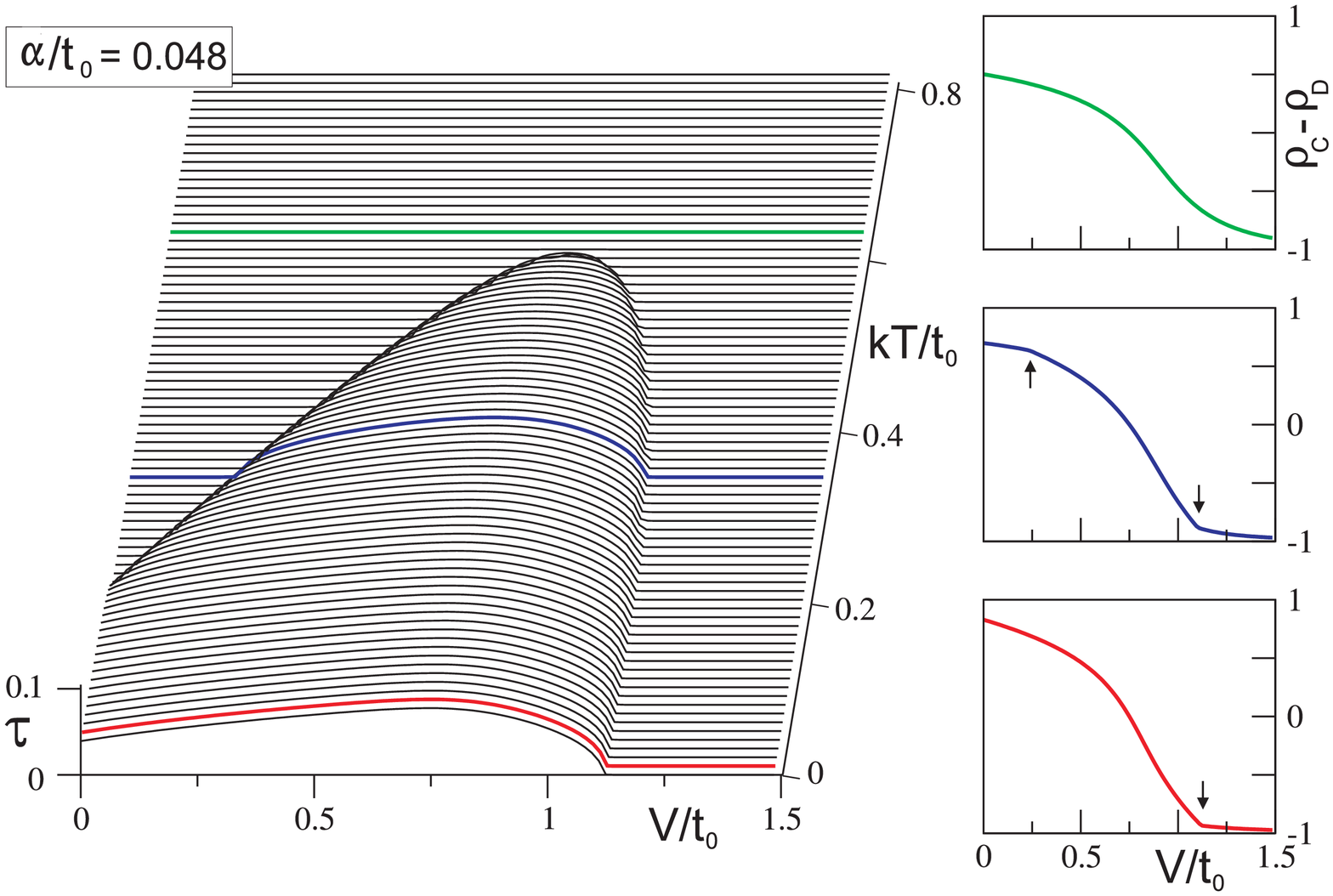}}}
\caption{(Color online) { \bf Left}: the CDW order parameter $\tau$ in the mean field treatment of equation \eqref{Hamiltonian}, as a function of temperature and the exciton binding energy (normalised with respect to the unperturbed bandwidth) for a constant value of the electron-hole coupling parameter. { \bf Right}: the charge transfer $\rho_C-\rho_D$ as a function of the exciton binding energy for selected temperatures, corresponding to the thick lines in the diagram on the left. The arrows indicate the transition to the ordered CDW state. Increasing the exciton binding energy leads to an increase in charge transfer, even in the absence of CDW order.}
\label{ChargeTransferFinal}
\end{figure}

Within the model system of equation \eqref{Hamiltonian}, it is clear that as a function of excitonic binding energy, there is a continuous interpolation between purely phononic CDW order and a purely Coulombic excitonic insulator state (see figure \ref{LineDiagramFinal}). For intermediate values of both the exciton binding energy and the electron-phonon coupling parameter, the presence of excitons can enhance the phonon-driven CDW order, resulting in an increase of the transition temperature with exciton binding energy, as in figure \ref{ChargeTransferFinal}. For temperatures above this transition, there is no CDW order and the mean field order parameter $\tau=\langle c^{\dagger}d+d^{\dagger}c \rangle$ is strictly zero. The exciton densities however, are not very strongly affected by this increase in temperature, as witnessed by the charge transfer diagrams on the right side of figure \ref{ChargeTransferFinal}. 

We can thus interpret the diffuse high temperature signal in the recent ARPES data of ref. \cite{Monney2} as arising from the presence of excitons above the CDW transition. These excitons are stable, bound particle-hole pairs which occur even in a mean-field treatment of the model of equation \eqref{Hamiltonian}. The presence of these excitons in TiSe$_2$ at high temperatures does not directly lead to the formation of a charge density wave because there is no nesting vector connecting the valence and conduction bands. They show up as a diffuse signal in the ARPES measurements due to higher-order processes involving these excitons with average total momentum close to the Brillouin zone boundary.

Once the temperature is lowered below 202 K, the renormalised electron-phonon coupling (due to the presence of the excitons) becomes strong enough to induce a CDW transition. The ordering is driven by a condensation of phonons rather than excitons, and the ordering wave vector is automatically commensurate with the lattice. The presence of a finite density of incoherent excitons is crucial for the renormalisation of the phonon spectrum, and gives rise to the large amount of charge transfer between the valence and conduction bands. The shape of the bands can be fitted with the same functions as those used in ref. \cite{Monney1}, with the gap arising from the renormalised electron-phonon coupling rather than the exciton binding energy.

To differentiate between the exciton condensation scenario and the combined exciton-phonon interpretation, one needs optical experiments probing not only the electronic density, but also the phase coherence of the excitons (either directly \cite{Littlewood2}, or by addressing a sliding mode). If one could also access the phonon dispersion it may be possible to get a more quantitative estimate of the extent to which excitons and phonons contribute to the CDW transition in TiSe$_2$ \cite{Wilson}, and consequently how large their role is in the formation of the recently discovered superconducting order at high pressure \cite{Anna,vanWezel}.

\subsection*{Conclusions}
We have shown that the recent high-resolution ARPES measurements on TiSe$_2$ can be interpreted in various different ways. Of these, the scenario introduced in ref. \cite{vanWezel}, which combines exciton formation with electron-phonon coupling, is particularly effective in explaining all the observed behaviours of this material. The diffuse signal of recent ARPES measurements above the transition temperature can be interpreted within a mean-field treatment of the combined scenario as a result of the presence of a finite density of excitons in TiSe$_2$, persisting to high temperatures. The transition itself is driven by the condensation of (Jahn-Teller) phonons, which are softened due to their interaction with the excitons. Both the specific observed, commensurate distortion pattern and the large amount of charge transfer between the valence and (backfolded) conduction bands can be naturally explained at the mean-field level as consequences of the involvement of hybridised phonons and excitons in the transition.

\subsection*{Acknowledgements}
The authors would like to thank the EPSRC and Jesus and Homerton Colleges of the University of Cambridge for financial support.

\end{document}